\documentclass[referee]{aa} 

\usepackage{graphicx,natbib}
\bibpunct{(}{)}{;}{a}{}{,}

\begin{document}

\sloppypar

\title{Thermal Timescale Mass Transfer Rates in Intermediate-Mass X-ray Binaries}

\author{Xiao-Jie Xu and Xiang-Dong Li}
\institute{Department of Astronomy, Nanjing University, Nanjing 210093, China\\
\email{lixd@nju.edu.cn} }
\date{}
\titlerunning{Thermal Timescale Mass Transfer Rates in IMXBs}
\authorrunning{Xu \& Li }

\abstract{Thermal timescale mass transfer generally occurs in close
binaries where the donor star is more massive than the accreting
star. The mass transfer rates are usually estimated in terms of the
Kelvin-Helmholtz timescale of the donor star. But recent
investigations indicate that this method may overestimate the real
mass transfer rates in accreting white dwarf or neutron star binary
systems. We have systematically investigated the thermal-timescale
mass transfer processes in intermediate-mass X-ray binaries, by
calculating binary evolution sequences with various initial donor
masses and orbital periods. From the calculated results we find that
on average the mass transfer rates are lower than traditional
estimates by a factor of $\sim 4$.

   }

   \maketitle

%

%
%
\section{Introduction}           
\label{sect:intro}

X-Ray binaries with neutron star accretors are traditionally divided
in to two groups based on the masses of the donor stars. One is
low-mass X-ray binaries (LMXBs) with donor stars less massive than
$\sim 1.5 M_\odot$, the other is high-mass X-ray binaries (HMXBs)
with donor masses exceeding $\sim 10.0M_\odot$. In LMXBs mass is
exchanged through Roche-Lobe overflow (RLOF), while in HMXBs, the
accretor is likely to be fed by the stellar wind-induced mass loss
of the companion. Systems with donor masses between 1.5 and
$10.0M_\odot$ are called intermediate-mass X-ray binaries (IMXBs).
Few IMXBs have been discovered in the Galaxy. The reason is that, on
one hand, mass transfer via RLOF is thought to be rapid and unstable
due to the large mass ratio, leading to the formation of a common
envelope; on the other hand, the donor stars at this mass range are
unable to generate strong winds to power bright X-ray emission from
the neutron star \citep{vdh75}.

Recent investigations on IMXB evolution lead to important
realization on the stability of super-Eddington mass transfer, and
suggest that many, or perhaps most, of the current LMXBs descended
from IMXBs. The studies on the evolution of the LMXB Cyg X-2
\citep{king99,pod00,kol00} indicate that the mass of the donor star
in this system must have been substantially larger ($\sim 3.5
M_{\sun}$) than its current value ($\sim 0.6 M_{\sun}$), implying
that intermediate-mass systems can survive the high mass transfer
phase by ejecting most of the transferred mass. The calculations by
\citet{tau00} have shown that the evolution of some IMXBs may
survive a spiral-in and experience a highly super-Eddington mass
transfer phase on a (sub)thermal timescale if the convective
envelope of the donor star is not too deep. These systems provide a
new formation channel for binary millisecond pulsars with heavy CO
white dwarfs and relatively short orbital periods ($3-50$ days).
\citet{dav98} have independently suggested that IMXBs may be the
progenitors of recycled pulsars in globular clusters. These works
emphasize the necessity of accurately defining an evolutionary path
of IMXBs, and motivate systematic analysis of binary systems
undergoing thermal timescale mass transfer.

Since the donor star in an IMXB is more massive than the accretor,
the RL radius of the donor will shrink during the mass transfer. At
the same time the donor star will either grow or shrink due to mass
loss. The stability of the mass transfer depends on the radius-mass
exponents for the donor and its RL \citep{sob97},
$\xi_{2}=(\partial\ln R_2/\partial\ln M_2)$ and
$\xi_{L}=(\partial\ln R_L/\partial\ln M_2)_{L}$, where $\xi_{2}$ is
the adiabatic or thermal response of the donor star to mass loss
($M_2$ and $R_2$ are the mass and radius of the donor star, $R_L$ is
its Roche-lobe radius, respectively). In general $R_L$ decreases
($\xi_{L} > 0$) when material is transfered from a relatively heavy
donor to a light accretor, and vice versa. Donor stars with
radiative envelopes will usually shrink ($\xi_{2} > 0$) in response
to mass loss, while donor stars with a deep convective envelope
expand rapidly ($\xi_{2} < 0$) in response to mass loss. The
relative sizes of these parameters determine whether the mass
transfer proceeds on either dynamical or thermal timescale. If
$\xi_{2}>\xi_{L}$ the mass transfer is dynamically stable, occurring
on either nuclear or thermal timescale. If $\xi_{2}<\xi_{L}$ the
Roche lobe radius shrinks more rapidly than the adiabatic radius,
and the mass transfer proceeds on dynamical timescale which leads to
a common envelope and a spiral in phase. The final product could be
either a Thorn-\.{Z}ytkow object or a short-period binary if the
envelope is ejected \citep[e.g.][]{tau00,pod02}.

\citet{pod02} made a survey of the X-ray binary sequences with the
donor masses ranging from 0.6 to $7 M_\odot$. These authors found
the actual mass transfer rates through RLOF sometimes deviate from
the values given by the traditional formula for thermal mass
transfer\footnote{In \citet{rap94} the thermal timescale mass
transfer rate is taken to be half of that given by Eq.~(1).},
\begin{equation}\label{eq:01}
\dot M_{\rm th}\simeq \frac{(M_{2}^{\rm i}-M_{1}^{\rm i})}{\tau_{\rm
KH}},
\end{equation}
where $M_1^{\rm i}$ and $M_2^{\rm i}$ are the initial masses of the
accretor and the donor respectively, and $\tau_{\rm KH}$ is the
Kelvin-Helmholtz time scale
\begin{equation}\label{eq:02}
\tau_{\rm KH}\simeq \frac{GM_2^2}{2R_2L_2},
\end{equation}
where $G$ is the gravitational constant, and $L_2$ the luminosity of
the donor star. The work done by Langer et al. (2000) on the
evolution of white dwarf binaries also indicates that Eq.~(1) could
overestimate the mean mass transfer rates by a factor of a few.

However, due to its simplicity, Eq.~(1) has been widely used in
population synthesis investigations \citep[e.g.][]{hur02,bel07} for
thermal timescale mass transfer in close binaries. The aim of this
paper is to present a modified empirical formula to estimate the
mean thermal timescale mass transfer rates onto neutron stars, by
calculating the evolutions of IMXBs systematically. The results may
be helpful to future investigations involving mass transfer
processes in IMXBs. We describe the stellar evolution code and the
binary model used in this study in \S 2. The calculated results and
fitting formulae for the mass transfer rates are presented in \S 3.
We conclude in \S4.

\section{Binary calculations}
\label{}

We have followed the evolution of the binary systems containing a
neutron star and an intermediate-mass secondary star using an
updated version of the evolution code developed by \citet[][see also
Pols et al. 1995]{egg71}. The opacities in the code are from
\citet{rog92} and from \citet{ale94} for temperatures below
$10^{3.8}$ K. We assume a mixing length parameter of $\alpha=2$, and
set the convective overshooting parameter to be 0.2. The metallicity
of the secondary is taken to be $Z=0.02$ and 0.001, and the
corresponding Helium abundance is 0.28 and 0.242, respectively. Each
system is set to start with a neutron star of mass $M_1=1.4M_\odot$
and a secondary of mass $M_2$ from 1.6 to $4.0 M_{\sun}$. Systems
with donor mass higher than $\sim 4.0 M_{\sun}$ always experience
unstable dynamical mass transfer \citep{tau00,pod02}. The effective
RL radius of the secondary is calculated with Eggleton's equation
\citep{egg83},
\begin{equation}
\frac{R_L}{a} =\frac {0.49 q^{2/3}}{0.6q^{2/3} + \ln(1+ q^{1/3})},
\end{equation}
where $a$ is the orbital separation, and $q=M_2/M_1$ is the mass
ratio. We use the following formula to calculate the mass transfer
rate from the donor star via RLOF \citep{egg71}
\begin{equation}
\dot{M_2}=-RMT(R_2-R_{L})^3,
\end{equation}
where  $RMT$ is a parameter adjusted automatically in the code,
usually taken to be 500. The mass loss of the secondary via stellar
wind is calculated according to the empirical formula given by
\citet{nie90},
\begin{equation}
\log(-\dot M_{2\rm w})= -14.02+
1.24\log(\frac{L_2}{L_{\odot}})+0.16\log(\frac{M_2}{M_{\odot}})+0.81\log(\frac{R_2}{R_{\odot}}).
\end{equation}
To follow the details of the mass transfer processes, we also
include losses of orbital angular momentum due to mass loss,
magnetic braking, and gravitational-wave radiation, although the
last process is not important in this analysis. For magnetic
braking, we use the standard angular momentum prescription suggested
by \citet{rap83}. The Eddington luminosity of the neutron star is
$L_{\rm Edd}=4\pi GM_1m_{\rm p}/\sigma_{\rm T}\simeq 1.3\times
10^{38}(M_1/M_{\sun})$ ergs$^{-1}$, where $m_{\rm p}$ and
$\sigma_{\rm T}$ are proton mass and the cross section of Thompson
scattering, respectively. We limit the maximum accretion rate of the
neutron star to the Eddington accretion rate $\dot{M}_{\rm
Edd}=R_1L_{\rm Edd}/GM_1$ ($\sim 1.5\times
10^{-8}\,M_{\sun}$yr$^{-1}$ for a $1.4M_{\sun}$ neutron star), i.e.,
$\dot{M}_1=-f\dot{M}_2$, where
\begin{equation}
f=\left\{ \begin{array}{ll} 1 & \textrm{if
$-\dot{M}_2<\dot{M}_{\rm Edd}$}\\
-\dot{M}_{\rm Edd}/\dot{M}_2 & \textrm{if $-\dot{M}_2 \ge
\dot{M}_{\rm Edd}$}
\end{array} \right..
 \label{eq:06}
\end{equation}
We let the excess mass be lost from the system with the specific
orbital angular momentum of the neutron star. The orbital separation
then changes according to the following equation
\citep[e.g.][]{sob97}
\begin{equation}\label{eq:19}
\frac {\dot a}{a}= \frac {2\dot J}{J}-\frac {2\dot M_2}{M_2}-\frac
{2\dot M_1}{M_1}+\frac {\dot M}{M}=\frac {2\dot{J}_{\rm
MB}}{J}-\frac {2\dot M_2}{M_2}[(1-q)+\frac {q(1-f)}{2(1+q)}]
\end{equation}
where $M=M_1+M_2$ is the total mass, $J$ the orbital angular
momentum, and $\dot{J}_{\rm MB}$ the rate of orbital angular
momentum loss by magnetic brakinig.

\section{Results}

We have calculated a large number evolutionary sequences for IMXBs
with various initial donor mass and orbital period, so that mass
transfer starts when the donor star is on early and late main
sequence (cases a1 and a2), in the Hertzprung gap (cases b1, b2, b3)
and on the giant branch (cases c1, c2 and c3), respectively. In
Fig.~1 we show the initial distribution of the binaries in the $M_2$
vs. $\log a$ diagram. Triangles, squares, and diamonds in the figure
correspond to the donor stars being main-sequence stars, subgiants,
and red giants at the onset of mass transfer, respectively. We stop
the calculation when either the mass transfer becomes dynamically
unstable or dominated by the nuclear evolution of the donor.
Obviously the occurrence of thermal timescale mass transfer depends
on the initial mass ratio and the orbital period. When $Z=0.02$ we
find that steady case a1 to case b2 thermal timescale mass transfer
is possible if the initial donor mass is between 1.6 and $3.6
M_\odot$, while systems either containing a more massive donor or in
cases b3 and c3 are subject to delayed unstable dynamical mass
transfer. When $Z=0.001$ the calculated results show that systems
with donors between 1.8 and $3.6 M_\odot$ experience stable thermal
timescale mass transfer in cases a1 to c2, otherwise the mass
transfer is dynamically unstable.

Typical examples of the evolutionary sequences with $Z=0.02$ and
$0.001$ are shown in Figs.~2 and 3 respectively. In Fig. 2 the
system contains a neutron star and a companion with initial mass of
$3.0 M_\odot$, which starts filling its RL roughly at the end of its
central hydrogen burning. In Fig. 3 the donor has an initial mass of
$3.6 M_\odot$ and starts filling its RL right after its helium
ignition. In the figures we demonstrate the evolution of the mass
transfer rate, the orbital period, the donor mass, and the neutron
star mass with time. In Fig.~2 the mass transfer rate first rises
rapidly to $\sim 10^{-5.5}\,M_{\odot}$yr$^{-1}$, then declines to a
few $10^{-8}\,M_{\odot}$yr$^{-1}$ after $\sim 9$ Myr, and stays
around this value for $\sim 1$ Myr. During the former rapid mass
transfer phase, the donor mass decreases from $3\,M_{\odot}$ to $<
1\,M_{\odot}$, but most of the mass is lost from the system, and
efficient accretion by the neutron star occurs during the latter
part of the mass transfer phase. The orbital period first decreases
to around 1.2 day, and then increases to $\sim 30$ day at the end of
mass transfer. Mass transfer shown in Fig.~3 is more rapid due to
the more massive and evolved donor star, lasting around 0.1 Myr.
About $2.6\,M_{\sun}$ mass is transferred from the donor star during
this phase, most of which is lost from the system, and the neutron
star mass hardly changes.


In our work the initiation ($t_{\rm i}$) and termination time
($t_{\rm f}$) of the thermal timescale mass transfer is assumed to
be once the mass transfer rate exceeds and declines to the
Eddington limit of the neutron star. The mean mass transfer rate
$\dot M_{\rm mean}$ is calculated from the following equation,
\begin{equation}\label{eq:10}
\dot M_{\rm mean}=\frac {M_2^{\rm i}-M_2^{\rm f}}{t_{\rm f}-t_{\rm
i}},
\end{equation}
where $M_2^{\rm i}$ and $M_2^{\rm f}$ are the donor mass at
$t=t_{\rm i}$ and $t_{\rm f}$, respectively (stellar wind mass loss
is negligible). In Tables 1 and 2 we list the calculated values of
$t_{\rm i}$, $t_{\rm f}$, $M_2^{\rm i}$, $M_2^{\rm f}$, $\dot M_{\rm
mean}$ and the maximum mass transfer rates $\dot M_{\rm max}$ for
evolutions with $Z=0.02$ and 0.001, respectively. For comparison, we
also list the expected values of $\dot M_{\rm th}$ calculated with
Eq.~(1). Figures 4 shows the calculated mean mass transfer rates as
a function of $(M_{2}^{\rm i}-M_{1}^{\rm i})/\tau_{\rm KH}$. A
linear fit can be obtained to be
\begin{equation}\label{eq:17}
\dot M_{\rm mean}\simeq 0.28 \frac{(M_{2}^{\rm i}-M_{1}^{\rm
i})}{\tau_{\rm KH}}
\end{equation}
for $Z=0.02$, and
\begin{equation}\label{eq:18}
\dot M_{\rm mean}\simeq 0.24 \frac{(M_{2}^{\rm i}-M_{1}^{\rm
i})}{\tau_{\rm KH}}
\end{equation}
for $Z=0.001$.

\section{Summary and discussion}
\label{sect:conclusion}

Our numerical calculations show that there are stable
super-Eddington thermal timescale mass transfer processes in IMXB
systems with donor mass between 1.6 and $3.6 M_\odot$ in both
$Z=0.02$ and $Z=0.001$ cases. We find that on average the
traditional expression (Eq.~[1]) overestimates the thermal timescale
mass transfer rates by a factor of $\sim 4$.

The results are obviously subject to various uncertainties in
treating the mass transfer processes in binary evolution. One of the
issues is the mass and angular momentum loss during mass transfer.
We have used Eddinton-limited accretion rate for neutron star
accretion. Recent observations of quite a few binary millisecond
radio pulsars constrain the pulsar masses to $\sim 1.35\,M_{\sun}$
\citep[][and references therein]{bas06}, suggesting that that almost
all of the transferred mass may be lost rather accreted by the
neutron star during the IMXB and LMXB phase. If the lost mass
carries the specific orbital angular momentum of the neutron star,
the orbital shrinking during thermal timescale mass transfer would
be slower than we have calculated. This can be clearly seen from
Eq.~(7) by setting $f=0$. The effect on mass transfer is most
significant for those binaries with mass transfer rates being mildly
supper-Eddington. For example, we find that the mean mass transfer
rate is decreased by a factor $\sim 5$ if the donor mass is $\sim
1.5-2.0\,M_{\sun}$ and $Z=0.02$. We note that in white dwarf binary
evolution, it has also been realized that the strong mass loss from
the accretor can stabilize the mass transfer even for a relatively
high mass ratio, avoiding the formation of a common envelope
\citep{hac96,li97,lan00}. If, however, part of the lost mass forms a
circumbinary disk rather leaves the system \citep{sob97}, the disk
would extract orbital angular momentum from the binary by tidal
torque, enhancing the mass transfer rates and leading to more rapid
orbital shrink \citep{spr01}.

It should be also noted that, during the evolution of an IMXB the
strong X-ray radiation by the accretor and the accretion disk could
illuminate the donor star and cause expansion of the donor and
strong stellar wind \citep{pod91,ham93,phi02}, which would also lead
to a higher mass transfer rate and shorter duration of mass
transfer. As there is not a generally accepted theory on the
irradiation effect, we have not include it in our calculations.

\begin{acknowledgements}
We are grateful to the referee for helpful comments. This work was
supported by the Natural Science Foundation of China under grant
numbers 10573010 and and 10221001.
\end{acknowledgements}

\clearpage
\begin{figure}
   \centering
   \resizebox{\hsize}{!}{\includegraphics{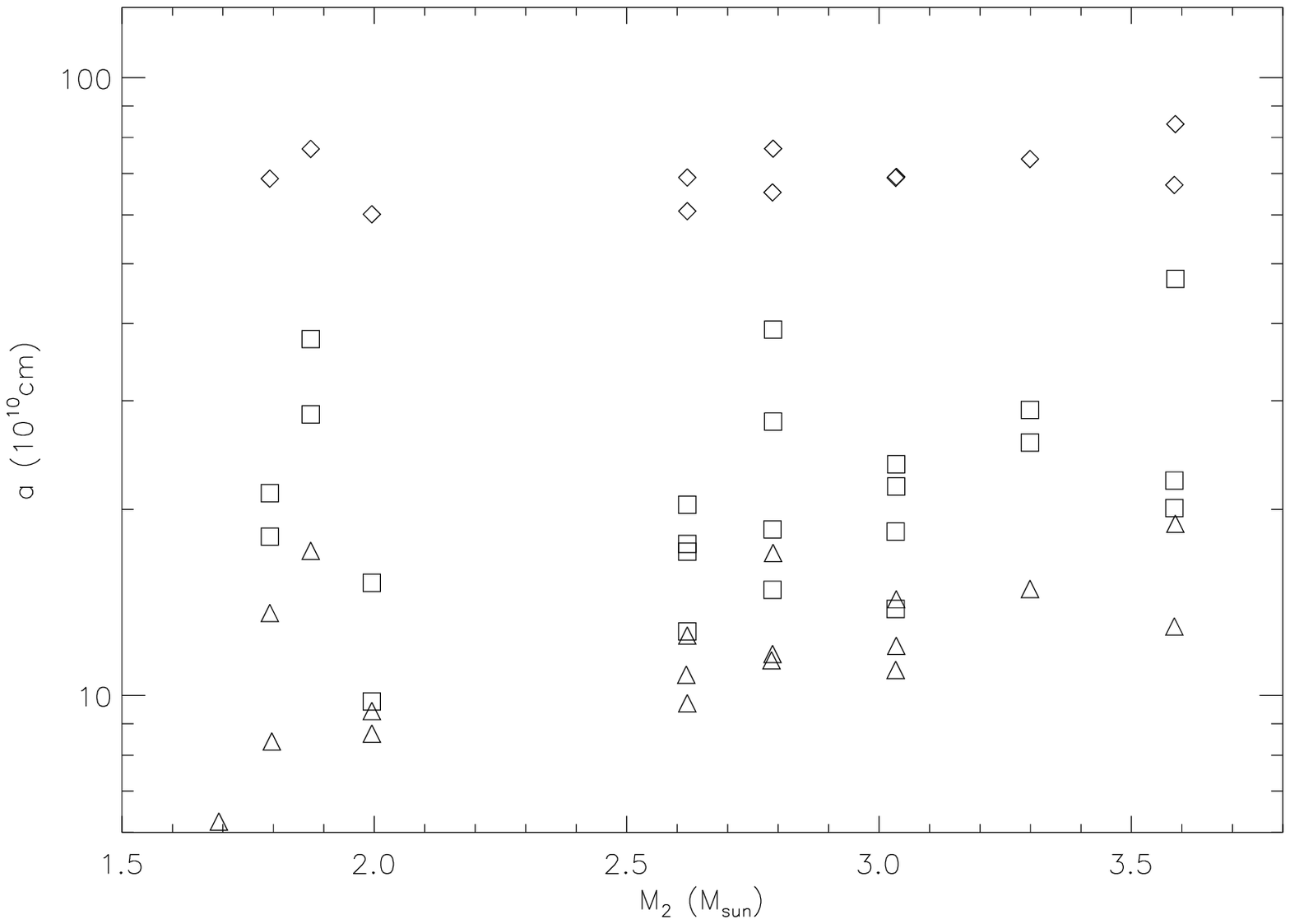}}
   \caption{The initial distribution of donor mass and orbital separation at the onset of mass
   transfer. The triangles, squares and diamonds refer to donors on main sequence,
     subgiant branch and red giant branch, respectively.}
   \label{Fig:01}
   \end{figure}

\clearpage
\begin{figure}
   \centering
   \resizebox{\hsize}{!}{\includegraphics{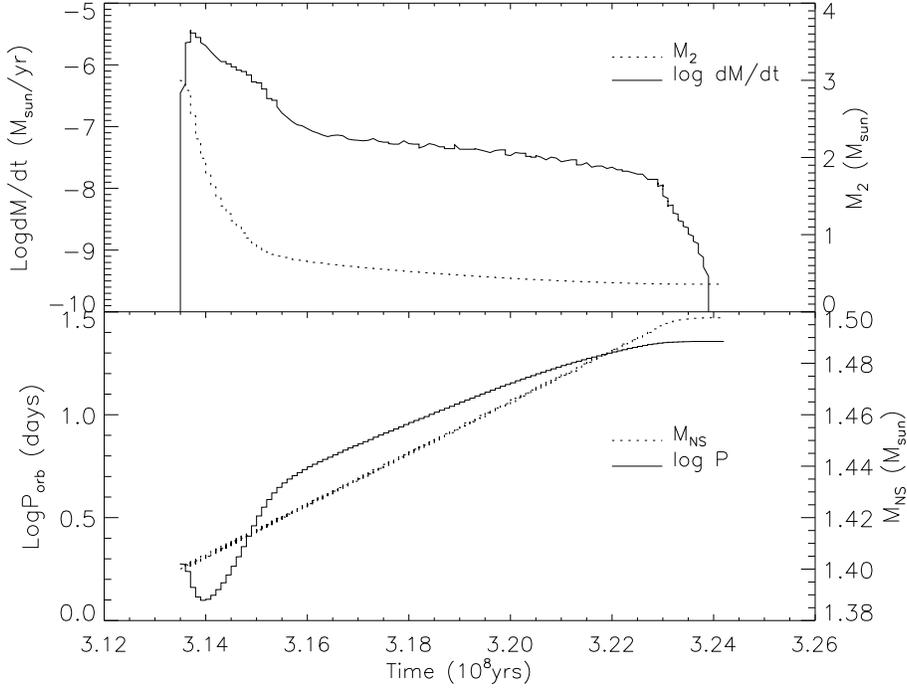}}
   \caption{The evolution of an IMXB with the donor of mass $3.0M_{\odot}$ and
   $Z=0.02$. The mass transfer starts right after the depletion of its core
   hydrogen. The solid and dotted lines  show the evolution
   of the mass transfer rate and the donor mass in the upper panel, and the evolution
   of the orbital period and the neutron star mass in the lower panel, respectively.}
   \label{Fig:02}
   \end{figure}

\clearpage
\begin{figure}
   \centering
   \resizebox{\hsize}{!}{\includegraphics{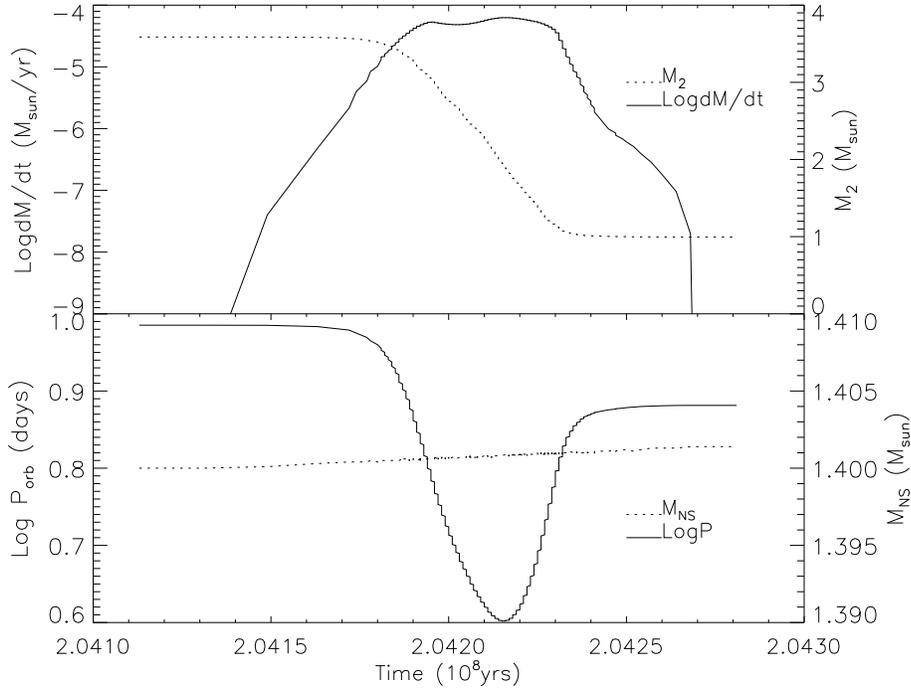}}
   \caption{Similar as Fig1. The evolution of an IMXB with the donor of mass $3.6M_{\odot}$ and
   $Z=0.001$. The mass transfer starts right after its central helium ignition .
   The solid and dotted lines  show the evolution
   of the mass transfer rate and the donor mass in the upper panel, and the evolution
   of the orbital period and the neutron star mass in the lower panel, respectively.
   }
   \label{Fig:03}
   \end{figure}


\clearpage
\begin{figure}
   \centering
   \resizebox{\hsize}{!}{\includegraphics{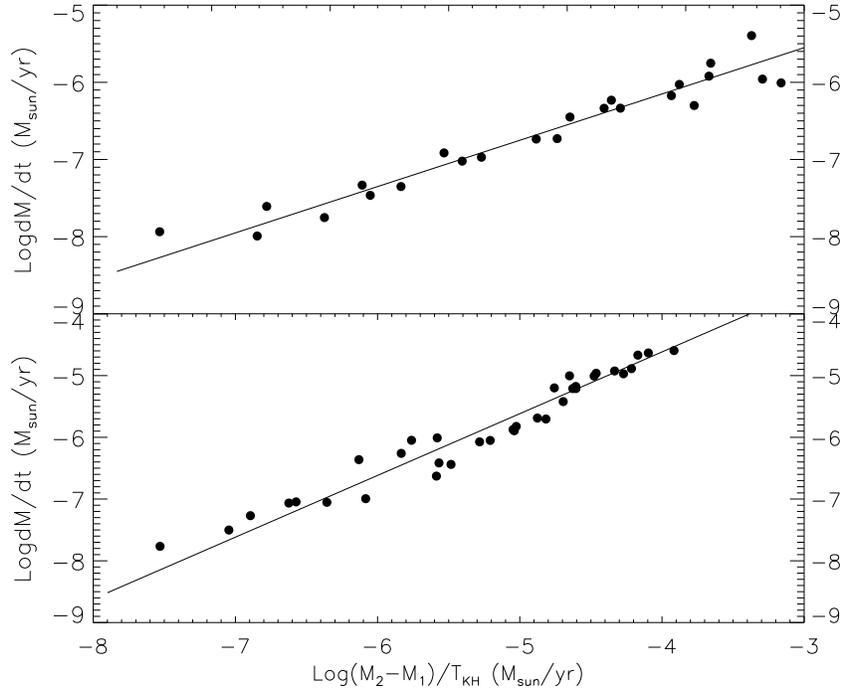}}
   \caption{The upper and lower panels show the fitting results for the mean thermal timescale mass transfer
   rates in $Z=0.02$ and $Z=0.001$ systems, respectively. The filled dots indicate the calculated results
   in this work.}
   \label{Fig:04}
   \end{figure}

\clearpage
\begin{table}[]
  \caption[]{The calculated data of 24 evolutionary sequences with stable
  thermal timescale mass transfer ($Z=0.2$). $\dot M_{\rm mean}$, $\dot M_{\rm max}$
  and $\dot M_{\rm th}$ are the mean mass transfer rates, the maximum mass transfer rates and
  the mass transfer rates calculated from Eq.~(1), respectively.}
  \label{Tab:highz}
  \begin{center}\begin{tabular}{ccccccccc}
  \hline\hline
$M_2$ & Case &  $t_{\rm i}$ & $M_{2}^{\rm i}$ & $t_{\rm f}$ &
$M_{2}^{\rm f}$ & $\log {\dot M_{\rm mean}}$ & $\log {\dot M_{\rm
max
}}$ & $\log {\dot M_{\rm th}}$  \\
($M_{\odot}$)& &($10^8$ yr)&($M_{\odot}$)&($10^8$
yr)&($M_{\odot}$)&($M_{\odot}$yr$^{-1}$)&($M_{\odot}$yr$^{-1}$)&($M_{\odot}$yr$^{-1}$)\\
  \hline\noalign{\smallskip}
1.6  & a1 & 7.52  &  1.58   &  7.58  &  1.51  &  -7.94  &   -7.90  & -7.72    \\
1.8  & a1 & 4.99  &  1.80   & 5.05   & 1.63   & -7.61   &   -7.55 &  -7.27   \\
2.0  & a1 & 5.10  &  2.00   &  5.18  &  1.61  & -7.33    & -7.14  & -6.87    \\
2.3  & a1 & 3.51  &   2.30  &  3.57  & 1.52   &  -6.91  &  -6.54 & -6.52    \\
3.0  & a1 & 1.56  &  3.00   & 1.61   & 1.26   &  -6.45   & -5.70  &  -5.99   \\
3.3  & a1 & 1.21  &   3.30  & 1.25   & 1.13   & -6.23   & -5.48  & -5.81    \\
1.8  & a2 & 11.0  &  1.80   & 11.1   & 1.64   & -7.75   &  -7.58 & -7.02     \\
2.0  & a2 & 8.06  &   2.00  &  8.15  & 1.60   & -7.35    & -6.97  & -6.70    \\
2.3  & a2 &  5.31 &   2.30  &  5.39  &  1.50  &  -6.97  &  -6.41 & -6.36    \\
2.6  & a2 & 4.09  &  2.61   &  4.16  &  1.36  & -6.73   & -5.99  & -6.04    \\
3.0  & a2 & 2.68  &  3.00   &  2.73  & 1.02   & -6.34   &  -5.56 & -5.78    \\
3.3  & a2 &  2.19 &  3.30   & 2.23   &  0.95 & -6.17   &  -5.35 & -5.56     \\
3.6  & a2 &  1.73 &  3.60   & 1.76   & 0.84  & -5.92   & -4.92   & -5.40    \\
1.6  & b1 & 18.8  & 1.57    & 18.8   & 1.52   & -7.99    & -7.96  & -7.31    \\
1.8  & b1 &  13.2 &  1.80   & 13.2   &  1.59  & -7.46   &  -7.29  & -6.83    \\
2.0  & b1 & 9.67  &  2.00   &  9.72  &  1.52  & -7.02  & -6.68   & -6.44    \\
 2.3 & b1 &  6.46 & 2.30    & 6.51   &  1.36  &  -6.7   & -6.20  & -6.13    \\
 2.6 & b1 &  4.61 &  2.60   & 4.64   &   1.03 &   -6.34 &  -5.83 & -5.84     \\
3.0  & b1 & 3.14  &  3.00   & 3.16   &  0.66  & -6.03   &  -5.44 & -5.53    \\
 3.3 & b1 & 2.42  &  3.30   &  2.43  &  0.65  &  -5.75  &  -5.22 &  -5.39    \\
 3.6 & b1 &  1.92 &  3.60   &  1.92  & 0.78  & -5.39  &  -4.91 & -5.22    \\
3.0  & b2 & 3.17  &  3.00   & 3.22   &  0.40 & -6.30    & -5.44  & -5.46    \\
 3.3 & b2 &  2.47 & 3.30    &  2.49  &  0.44 & -5.96   &  -4.98 & -5.18    \\
 3.6 & b2 & 1.95  &  3.60   &  1.98  &  0.46 & -6.00    & -4.96  & -5.10    \\
\hline
\end{tabular}\end{center}
\end{table}

\clearpage
\begin{table}[]
  \caption[]{The calculated data of 34 evolutionary sequences with stable
  thermal timescale mass transfer ($Z=0.001$).}
  \label{Tab:lowz}
  \begin{center}\begin{tabular}{ccccccccc}
  \hline\hline
$M_2$ & Case &  $t_{\rm i}$ & $M_{2}^{\rm i}$ & $t_{\rm f}$ &
$M_{2}^{\rm f}$ & $\log {\dot M_{\rm mean}}$  & $\log {\dot M_{\rm
max}}$& $\log \dot{M}_{\rm th}$  \\
($M_{\odot}$)& &($10^8$ yr)&($M_{\odot}$)&($10^8$
yr)&($M_{\odot}$)&($M_{\odot}$yr$^{-1}$)&($M_{\odot}$yr$^{-1}$)&($M_{\odot}$yr$^{-1}$)\\
  \hline\noalign{\smallskip}
1.8&a1&3.29&1.70&3.47&1.39&-7.77&  -7.48&-7.53\\
1.8&a2&7.33&1.80&7.52&1.21&-7.50&  -6.23&-7.05\\
1.8&b2&9.01&1.80&9.13&0.80&-7.05&  -6.21&-6.36\\
2.0&a2&5.52&2.00&5.67&1.17&-7.27&  -6.40&-6.89\\
2.0&b1&6.40&2.00&6.55&0.73&-7.05& -6.043 &-6.57\\
2.0&b2&6.85&1.87&6.96&0.69&-6.99& -4.14 &-6.08\\
2.3&c1&6.04&2.31&6.05&0.87&-5.69& -5.40 &-4.88\\
2.3&c2&5.75&2.31&5.76&0.96&-5.82&  -5.64&-5.03\\
2.3&c3&6.10&2.31&6.11&0.92&-5.70& -5.37 &-4.81\\
2.6&a2&3.07&2.62&3.11&0.84&-6.36& -5.50 &-6.13\\
2.6&b1&3.35&2.62&3.38&0.56&-6.05& -5.43 &-5.76\\
2.6&b2&3.42&2.62&3.44&0.47&-6.01& -5.31 &-5.58\\
2.6&c1&4.44&2.62&4.442&0.77&-5.21& -5.07 &-4.61\\
2.6&c2&4.31&2.62&4.31&0.73&-5.20&-4.91 &-4.76\\
2.6&c3&4.43&2.62&4.43&0.77&-5.21& -4.82&-4.63\\
2.8&a2&2.64&2.79&2.68&0.81&-6.26& -5.39 &-5.83\\
2.8&b1&2.93&2.79&2.99&0.40&-6.41&  -5.23&-5.57\\
2.8&b2&2.97&2.79&3.00&0.42&-6.07& -5.10 &-5.28\\
2.8&c1&3.80&2.78&3.80&0.82&-5.01&  -4.64&-4.48\\
2.8&c2&3.72&2.79&3.72&0.79&-5.18& -4.71 &-4.61\\
2.8&c3&3.76&2.79&3.76&0.81&-5.00& -4.67 &-4.65\\
3.0&a2&2.37&3.03&2.48&0.40&-6.63& -5.13 &-5.59\\
3.0&b1&2.40&3.03&2.47&0.41&-6.44&  -5.08&-5.48\\
3.0&b2&2.45&3.03&2.48&0.45&-6.05&  -4.94&-5.21\\
3.0&c1&3.10&3.03&3.11&0.89&-4.97& -4.42 &-4.27\\
3.0&c2&3.01&3.03&3.01&0.85&-4.96& -4.54 &-4.46\\
3.0&c3&3.11&3.03&3.12&0.89&-4.67&  -4.38&-4.17\\
3.3&b2&2.03&3.30&2.05&0.49&-5.87& -4.58 &-5.05\\
3.3&c1&2.54&3.29&2.55&0.97&-4.63&  -4.24&-4.10\\
3.3&c2&2.47&3.30&2.48&0.92&-4.93& -4.38 &-4.33\\
3.6&b1&1.69&3.59&1.71&0.52&-5.89& -4.05 &-5.04\\
3.6&b2&1.70&3.59&1.71&0.56&-5.42&  -3.94&-4.70\\
3.6&c1&2.10&3.59&2.10&1.05&-4.60&  -4.05&-3.92\\
3.6&c2&2.04&3.59&2.05&1.00&-4.89&  -4.20&-4.22\\
\hline
\end{tabular}\end{center}
\end{table}

\end{document}